\documentstyle[aps,prl]{revtex}  			
\begin{document}
\twocolumn[\hsize\textwidth\columnwidth\hsize\csname  	 
@twocolumnfalse\endcsname				
\draft
\title{Discrete-Lattice Model for Surface Bound States and Tunneling in d-Wave 
Superconductors}
\author{M. B. Walker and P. Pairor }
\address{Department of Physics, 
University of Toronto,
Toronto, Ont. M5S 1A7 }
\date{\today }
\maketitle

\widetext					%
\begin{abstract}
Surface bound states in a discrete-lattice model of a $d_{x^2 - y^2}$ cuprate superconductor are shown to be, in general, coherent superpositions of an incoming excitation and more than one outgoing excitation, and a simple graphical construction based on a surface Brillouin zone is developed to describe their nature.  In addition, a momentum-dependent lifetime contribution to the width of these bound states as observed in tunneling experiments is derived and elucidated in physical terms.
\end{abstract}


\vfill		
\narrowtext			%

\vskip2pc]	


One of the consequences of the change of sign of the gap function in the $d_{x^2 - y^2}$ model of the high-temperature cuprate superconductors is the existence of zero-energy surface bound states  which are believed to be responsible \protect\cite{hu97,tan95,mat95,buc95,xu96,fog97,zhu98} for the zero-bias conductance peaks (ZBCP's) observed \protect\cite{gee88,les92,kas94,cov96,alf97,eki97,wei98} in the in-plane conductance of tunnel junctions.  Further support for the idea that surface bound states are responsible for the ZBCP's comes from the study of the splitting of the ZBCP's in a magnetic field \protect\cite{fog97,cov96}.  Furthermore, the splitting of the ZBCP's observed \protect\cite{cov96} at low temperatures in zero magnetic field provides evidence \protect\cite{fog97,zhu98} for the existence of the so-called $d + is$ surface state \protect\cite{mat95,buc95,sig95} (where the order parameter acquires an $s$-wave component at certain surfaces which is $\pi/2$ out of phase with a suppressed $d$-wave component).

In current models \protect\cite{hu97,tan95,mat95,buc95,xu96,fog97,zhu98}, the surface bound state is a coherent combination of an incoming excitation and an outgoing excitation whose existence is conditional on the gaps of the two excitations having opposite signs. Since, in a $d_{x^2 - y^2}$ superconductor, the energy gap has changes of sign imposed by symmetry, surface bound states are a natural consequence of $d_{x^2-y^2}$ symmetry, which is one of the reasons for their current interest.

Here we study a discrete-lattice  model which gives surface bound states with new features.  In particular there are surface bound states which are coherent combinations of an incoming excitation with more than one outgoing excitation provided all outgoing excitations have gaps of the same sign, which must be different from the sign of the gap of the incoming excitation.  The existence of several conditions (one for each outgoing excitation) reduces the number of surface bound states, as can be seen below in detailed results for (210) and (110) surfaces. An important aspect of the work below is the introduction of a simple graphical construction, based on the application of Bloch's theorem to translations parallel to the surface and the idea of a surface Brillouin zone, which elucidates the more complex structure of the surface bound states in a discrete lattice model.

We also develop a theoretical expression for the ZBCP contribution to the tunneling conductance, and in particular elucidate a mechanism of a momentum-dependent broadening  of the surface bound states which must inevitably accompany attempts to observe these states in tunneling experiments.  This lifetime broadening comes about because a quasiparticle in a surface bound state on the superconducting side of a normal to insulating to superconductor (NIS) tunnel junction will ultimately tunnel into the normal metal.  If the probability of tunneling through the insulating barrier is $P({\bf k_{\parallel}})$ (which depends on the momentum parallel to the surface, ${\bf k_{\parallel}}$) then the lifetime $\tau ({\bf k_{\parallel}})$ will be given by $\tau^{-1}({\bf k_{\parallel}}) = P({\bf k_{\parallel}}) f({\bf k_{\parallel}})$ where $f({\bf k_{\parallel}})$ is the so-called attempt frequency, i.e. the frequency with which the quasiparticle returns to the surface.  (This is a familiar result from textbook studies of the alpha decay of nuclei.)  To calculate the attempt frequency, recall from previous studies \protect\cite{hu97,tan95,mat95,buc95,xu96,fog97,zhu98} of surface bound states that the distance the (exponentially decaying) bound state wave function extends into the superconductor is $\xi({\bf k_{\parallel}}) = \hbar v_{\perp}({\bf k_{\parallel}})/\Delta({\bf k_{\parallel}})$ where $\Delta$ is the gap and $v_{\perp}$ is the quasiparticle velocity normal to the surface; the attempt frequency is thus $f\sim v_{\perp}/(2\xi) = \Delta/(2\hbar).$  Thus, the energy width of the surface bound state is $\Gamma \sim \hbar /\tau$, i.e.
\begin{equation}
	\Gamma({\bf k_{\parallel}})=\case{1}{2} P({\bf k_{\parallel}}) 	\Delta({\bf k_{\parallel}}).
	\label{a}
\end{equation}
This result has a significant dependence on ${\bf k_{\parallel}}$ and is the width which will be relevant for sufficiently pure superconductors having sufficiently flat surfaces.  

Other articles have taken into account the broadening of the surface bound states through impurity scattering by adding a momentum-independent imaginary part to the energy \protect\cite{fog97,alf97}, and the broadening effect of surface roughness has also been considered \protect\cite{fog97}. Also, the lifetime effect elucidated here is presumably implicitly present in the numerical plots of tunneling conductance presented by some other authors \protect\cite{tan95,xu96}, but its physical origin as described above does not appear to have been commented upon, nor have explicit expressions for this lifetime and its momentum dependence been given.

For an NIN junction, we find for the conductance per surface unit cell, G$_N$, the result
\begin{equation}
	G_N = 2(e^2/h) \langle P({\bf k_{\parallel}})\rangle
	\label{b}
\end{equation}
where the angular brackets indicate an average over the wave vectors ${\bf k_{\parallel}}$.  This has the familiar and intuitively appealing form of a Landauer formula \protect\cite{lan81}.  Clearly a measurement of $G_N$ provides information relevant to an estimate of the width $\Gamma({\bf k_{\parallel}})$ through Eq.\ (\ref{a}).    

In the case of an NIS junction, we show below that the conductance per surface unit cell associated with the surface bound states, and which gives rise to a zero bias conductance peak (ZBCP), is
\begin{equation}
	G_{ZBCP} = 4\frac{e^2}{h} \left\langle \frac{\Gamma^2({\bf 	k_{\parallel}})}{(eV-E_B({\bf k_{\parallel}}))^2  
	 + \Gamma^2({\bf k_{\parallel}})} \right\rangle
	\label{c}
\end{equation}
where $\Gamma({\bf k_{\parallel}})$ is given by Eq.\ (\ref{a}), and $E_B({\bf k_{\parallel}})$ is the energy of the bound state.  The term in the angular brackets is the probability of Andreev reflection (clearly a resonant process) of an electron incident on the junction from the normal side [and having energy less than $\Delta({\bf k_{\parallel}})$].  Interestingly, when the voltage satisfies $eV=E_B({\bf k_{\parallel}})$, the probablilty of Andreev reflection for an electron at that value of ${\bf k_{\parallel}}$ is unity.  The factor of $4$ in this equation comes from one factor of $2$ accounting for both electron and hole processes, and a second factor of $2$ coming from the fact that the Andreev reflection of a single electron results in two electrons being transferred across the barrier into the superconductor.  Also, since $E_B({\bf k_{\parallel}})$ is an odd function of ${\bf k_{\parallel}}$, G$_{ZBCP}$ is an even function of the voltage.    While Eq.\ (\ref{c}) is limited in that it ignores the conductance associated with the usual propagating excitations in the superconductor, it is useful in that it gives a simple, intuitive and detailed expression for the ZBCP, which is a separate and clearly identifiable part of experimental conductance versus voltage data.  Also, although Eq.\ (\ref{c}) is derived below in detail only for the case of a (110) surface, results of the same form are obtained for different models of the surface with different results for $E_B$, and it is therefore tempting to speculate that the result is of general validity.

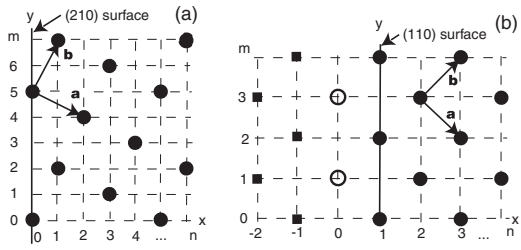
\begin{figure}
\caption{A (210) surface (a) and a (110) NIS tunnel junction (b).  In (b) the solid (open) circles, and the solid squares represent ions in the superconductor (insulator), and the normal metal, respectively.  The ionic columns and rows are labelled by n and by m.}
\label{fig1}
\end{figure}

We now outline some of the technical details, beginning with the description of the possible surface bound states for 
the (210) surface shown in Fig.\ \ref{fig1}(a).  We describe the 
superconducting state by the discrete-lattice Bogoliubov-de Gennes 
equations
\begin{equation}
\sum_{\bf x^\prime} \left[
		\begin{array}{rr}
			-t({\bf x, x^\prime}) & \Delta({\bf x, x^\prime}) \\
			\Delta^\ast({\bf x, x^\prime}) & t({\bf x, x^\prime})
		\end{array} \right]
	U({\bf x^\prime}) = E U({\bf x})
\label{1}
\end{equation}
where $U({\bf x})$ is the two-component wave function and ${\bf 
x}$ labels the ion positions.  For simplicity we consider nearest-
neighbor hopping only [$t({\bf x, x+a}) = t({\bf x, x+b}) = t$] and a $d$-wave order parameter [$\Delta ({\bf x, x+a}) = -\Delta({\bf x, x+b}) = \Delta$].  

Note that the crystal, including its surface, is invariant with respect 
to a translation parallel to the surface through a distance $\sqrt{5}a$.  Thus, from Bloch's theorem, the eigenstates of Eq.\ (\ref{1}) can be written in the form 
\begin{equation} 
	U({\bf x}) = e^{imk_y a/\sqrt{5}} U^{k_y}(n)
	\label{2}
\end{equation}
where $|k_y| < \pi/(\sqrt{5} a)$, and $m$ and $n$ are integer ionic 
row and column indices as indicated in Fig.\ \ref{fig1}.  Substituting Eq.\ (\ref{2}) 
into Eq.\ (\ref{1}) yields a one-dimensional equation
\begin{equation}
	\sum_{n^\prime} T^{k_y}(n,n^{\prime})U^{k_y}(n^{\prime})
	= E U^{k_y}(n)
	\label{3}
\end{equation}
for each $k_y$.  Note that the nearest-neighbor interaction between 
the ions in the two dimensional lattice couples the ionic column $n$ 
to columns $n^{\prime} = n-2, n-1, n, n+1$ and $n+2.$  Thus Eq.\ (\ref{3})
is a fourth order difference equation which will have four linearly 
independent solutions.

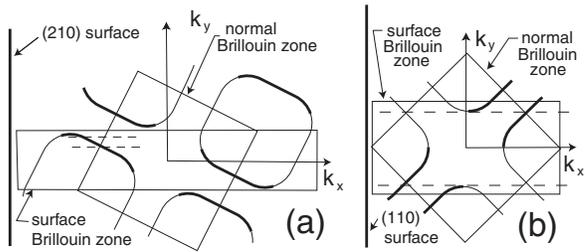
\begin{figure}
\caption{The surface Brillouin zone for a (210) surface (a), and a (110) surface (b).  The thick- and thin-line portions of the Fermi surface (sections of rounded squares) correspond to positive and negative gap functions, respectively.}
\label{fig2}
\end{figure}

Before describing the solutions of Eq.\ (\ref{3}) it is necessary to 
introduce the idea of a surface Brillouin zone.  The normal Brillouin 
zone for the square lattice of Fig.\ \ref{fig1} is  the square in Fig.\ \ref{fig2} containing the set of wavevectors $|k_a|, |k_b| < \pi /a$ .   The surface Brillouin zone is the rectangle in Fig.\ \ref{fig2}(a) containing the set of wave vectors $|k_x| < \sqrt{5} \pi /a, |k_y| < \pi / (\sqrt{5}a)$.  These two  Brillouin zones are equivalent in the sense that a complete set of bulk states can be equivalently described as having wave vectors in either Brillouin zone.  However, from Eq.\ (\ref{2}), an  arbitrary surface eigenstate is most simply described as a linear combination of bulk states whose wave vectors lie on a line of fixed $k_y$ in the surface Brillouin zone.  The Fermi surfaces in Fig.\ \ref{fig2} are drawn not as calculated from our model, but as determined from angular resolved photoemission studies \protect\cite{sch97,din96} from YBa$_2$Cu$_3$O$_{6+x}$ and Bi$_2$Sr$_2$CaCu$_2$O$_{8+x}$. 

The bulk solutions that can participate in the formation of a surface 
bound state with a given $k_y$ have energies less than the minimum 
energy gap $\Delta({\bf k})$ where ${\bf k}$ is one of the wave 
vectors where a line of fixed $k_y$ cuts the Fermi surface.  As in 
previous work \protect\cite{hu97,tan95,mat95,buc95,xu96,fog97,zhu98} such solutions can be found as an expansion in 
powers of $\Delta /\epsilon_F$, $\epsilon_F$ being the Fermi energy, and,  to lowest order in $\Delta$, are
\begin{equation}
	U^{k_y}_i(n) = \left[ 
			\begin{array}{c}
				\Delta_i \\
				E \pm i\Omega_i	
			\end{array} \right]
		e^{ink_{ix}a/\sqrt{5}} e^{-\kappa_i n}
	\label{4}							 
\end{equation}
Here ${\bf k_i}= (k_y,k_{ix})$ must be on the Fermi surface; also 
$\Delta_i = \Delta({\bf k_i})$, $\Omega_i = \sqrt{\Delta_i^2 - E^2}$, 
and $\kappa_i  = \Omega_i a/(\sqrt{5} \hbar |v_{ix}|)$ with $v_{ix}$
being the $x$ component of the normal-state electron velocity.  The 
upper and lower signs in Eq.\ (\ref{4}) correspond to 
$ v_{ix} < 0 $ and $v_{ix} > 0$,
respectively.  From Fig.\ \ref{fig2}(a) it is seen that for many values 
of $k_y$, a line of constant $k_y$ intersects the Fermi surface at four 
distinct points, giving four values $k_{ix}$, i= 1 ... 4, and hence the 
required four linearly independent solutions.
In these cases the surface bound state solutions will have the form
\begin{equation}
	U^{k_y}(n) = \Sigma_i C_i U^{k_y}_i(n),		
	\label{5}
\end{equation}
where only one incoming wave (i.e. $v_{1x} < 0$) is included, for which 
$C_1$ is taken to be unity, as are two outgoing waves for which $C_2$ and $C_3$ (in general of comparable magnitude) are to be determined.  
The boundary conditions determining $C_2$, $C_3$ and the surface bound state energy $E$ are Eqs.\ (\ref{3})
for $n=0$ and $n=1$, which
yield
\begin{equation}
	U^{k_y}(-1) = U^{k_y}(-2) = 0.			
	\label{6}
\end{equation}
A surface bound state with $E=0$ is found if the outgoing wave gaps $\Delta_2$ and $\Delta_3$ 
have the same sign, and are opposite in sign to the incoming 
wave gap $\Delta_1$.

If a line of constant $k_y$ intersects the Fermi surface at only two 
points in the surface Brillouin zone, only two solutions (say with gaps $\Delta_1$ and $\Delta_2$) are of the 
form of Eq.\ (\ref{4}), and a further two solutions must be sought 
numerically.  As in Eq.\ (\ref{4}), these solutions are proportional to 
$exp(iqn)$ where $q$ is complex.  In Eq.\ (\ref{4}), however, $Im(q) \ll 1$
by virtue of the approximation $(\Delta_i/\epsilon_F) \ll 1$, whereas 
in the present case the two additional solutions will have $Im(q) \sim 1$ and are thus localized close to the surface.  All four solutions are included in Eq.\ (\ref{5}), which gives a surface 
bound state with $E=0$ provided $\Delta_1 \Delta_2 < 0$.

Applying the conditions just established shows that the only surface bound states  which can occur on a (210) surface are those made up from incoming waves whose wave vectors fall on the Fermi between the two dashed lines in Fig.\ \ref{fig2}.  There are relatively few of these bound states because having to satisfy a condition on each of two outgoing waves is more restrictive than having to satisfy a condition on only one outgoing wave, as in other studies \protect\cite{hu97,tan95,mat95,buc95,xu96,fog97}.

We have  also determined the surface bound states for a (110) surface (as in Fig.\ \ref{fig1}(b) but including only ion columns $n \geq 1$), and with the model extended to include a second-nearest-neighbor hopping  interaction $t({\bf x, x+a+b})$ plus related terms derived from the tetragonal symmetry (which gives a more realistic Fermi surface, comparable to that shown in Fig.\ \ref{fig2}).  
The formal aspects of the problem, including the boundary condition and the conditions for the existence of surface bound states, are analogous to the problem studied above of a (210) surface for the case of nearest-neighbor interactions.  Applying these conditions leads to the conclusion that surface bound states with zero energy exist only for those values of $k_y$ which lie between the two dashed lines in the surface Brillouin zone in Fig.\ \ref{fig2}(b).

We now describe the calculation of the ZBCP tunneling conductance for the NIS junction at the (110) surface shown in Fig.\ \ref{fig1}(b), restricting our discussion to the features particular to our model, since the general features of such calculations are well established \protect\cite{tan95,xu96,zhu98}.  The ions of the column $n=0$ in Fig.\ \ref{fig1}(b) have an additional repulsive potential $v > 0$ and thus represent an insulating barrier.  We return, for simplicity, to the case of a nearest-neighbor only hopping interaction, $t$ (assumed to be the same for all pairs, metal, insulator, or superconductor) giving a normal-state quasiparticle energy $\epsilon({\bf k}) = -4t cos(q_x)cos(q_y)$ where $q_x,_y = k_x,_y a/\sqrt{2}.$

Assuming for the moment that the metals on both sides of the insulating barrier are normal, the probability of transmission of
an electron on the Fermi surface across the barrier is easily found to be
\begin{equation}
	P(k_y) = 16(t/v)^2 (cos^2q_y - \nu^2).
	\label{7}
\end{equation}
The chemical potential has been written in terms of $\nu$ by $\mu = 4t\nu$ (so that $\nu = 0$ and $\nu = 1$ correspond to half-filled and full bands, respectively) and $q_x$ is determined in terms of $q_y$ for electrons on the Fermi surface by $cos q_x = -\nu/cos q_y.$  As expected, it is easiest for electrons incident normally on the barrier to be transmitted, and the probability of transmission goes to zero for electrons travelling parallel to the barrier.

As described above, a constant nearest-neighbor gap is assumed for ions in the superconductor except at the surface where we add an additional gap $\delta \Delta({\bf x, x+a}) \equiv \Delta_a$ and $\delta \Delta({\bf x, x+b}) \equiv \Delta_b$; this additional gap is nonzero only for ${\bf x}$ lying on the superconducting surface, i.e on the column $n = 1$.  The
definitions of $\Delta_s$ and$\Delta_d$ (both real) by $\Delta_a + \Delta_b = 2i\Delta_s$ and $\Delta_a-\Delta_b = 2\Delta_d$ corresponds to the existence of a $d+is$ state at the surface, as suggested in \protect\cite{mat95,buc95,fog97,zhu98,sig95}.  Also, although our calculation is not self-consistent, the expected suppression of the d-wave component of the gap at the surface can be mimicked by an appropriate choice of $\Delta_d$.

Making use of the experimental fact that the width of the ZBCP is much less than $\Delta$, and also taking $E \ll \Delta$, we find that the tunneling conductance is given by Eq.\ (\ref{c}),  where the probability of Andreev reflection of an electron incident on the junction from the normal side is given by the expression in angular brackets in this equation.  Also, $\Gamma(k_y)$ is given by Eq.\ (\ref{a}) with $P(k_y)$ given by Eq.\ (\ref{7}) and $\Delta(k_y) = 4\Delta (cos^2 q_y - \nu^2)^{1/2} tan q_y.$  The energy of the surface bound state, which is independent of $\Delta_d$, i.e. of the d-wave gap suppression at the surface, is given by
\begin{equation}
	E_B = -32\nu \Delta (\Delta_s/t) [1 - (\nu^2/ cos^2 q_y)] tan 			q_y.
	\label{10}
\end{equation}

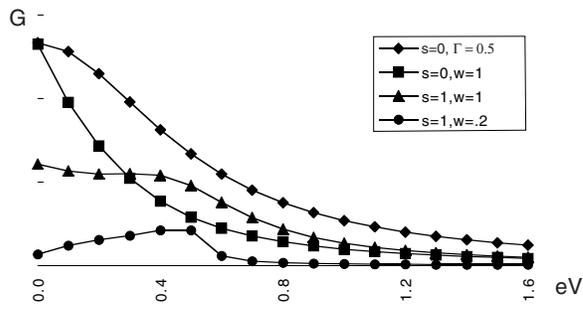
\begin{figure}
\caption{The ZBCP for a (110) NIS junction as a function of the shift s and the width w of the surface bound state.}
\label{fig3}
\end{figure}

Fig.\ \ref{fig3} shows the tunneling conductance as evaluated from Eq.\ (\ref{c}) for several values of the parameters $w = 16 \Delta (t/v)^2$ and $s = 16 \Delta (\Delta_s/t)$ giving the magnitudes of the bound state width and shift, respectively.  (Note that so long as $w, s$ and $eV$ have the same units, these units are irrelevant.)  In comparing the two curves for which there is no surface $s+id$ state, i.e. $s=0$, note that the momentum dependence of $\Gamma$ in the $w=1$ case gives a sharper cusp-like behavior at the origin  relative to the case $\Gamma = 0.5$ and independent of momentum.  There is also a weak cusp-like behavior at the origin in the broad $s=w=1$ curve, reflecting the fact that the low energy bound states have a low attempt frequency and are thus relatively narrow in energy. (Note that zero energy peaks in a broad ZBCP have been observed \protect\cite{cov96}.) Also, when $w \ll s$, the conductance gradually increases when $V$ is increased from zero, and then drops rapidly to zero for $eV$ greater than the maximum bound state energy.

In conclusion, we note that the analysis of a discrete-lattice model of surface bound states has uncovered new behavior, namely bound states made up of coherent superpositions of multiple excitations, and that a simple graphical approach based on the idea of a surface Brillouin zone helps to sort out the detailed structure of the bound states.  Furthermore a qualitative ``alpha decay'' type of picture (backed up by detailed calculation) is given of the momentum-dependent width of the surface bound states which would be observed in sufficiently pure superconductors with sufficiently flat surfaces.  Finally, a formula [Eq.\ (\ref{c})] for the ZBCP conductance is given which has the possibility to be general in applicability.

Helpful discussions with M.E. Zhitomirsky are acknowledged, as is the support of the Natural Sciences and Engineering Research Council of Canada.

\end{document}